\documentclass[reprint,amsmath,amssymb,aps,pra, superscriptaddress]{revtex4-2}

\usepackage{hyperref}
\usepackage{graphicx}
\usepackage{bm,color}
\usepackage{braket}
\usepackage{siunitx}
\usepackage{float}

\newcommand{\e}{\mathrm{e}}
\renewcommand{\d}{\mathrm{d}}
\renewcommand{\i}{\mathrm{i}}

\begin{document}

\bibliographystyle{apsrev4-2}

\title{Gate-based protocol simulations for quantum repeaters using quantum-dot molecules in switchable electric fields}

\author{Steffen Wilksen}
\affiliation{Institute for Theoretical Physics and Bremen Center for Computational Material Science, University of Bremen, Bremen, Germany}

\author{Frederik Lohof}
\affiliation{Institute for Theoretical Physics and Bremen Center for Computational Material Science, University of Bremen, Bremen, Germany}

\author{Isabell Willmann}
\affiliation{Institute for Theoretical Physics and Bremen Center for Computational Material Science, University of Bremen, Bremen, Germany}

\author{Frederik Bopp}
\affiliation{Walter Schottky Institut, School of Natural Sciences, and MCQST, Technische Universität München, Am Coulombwall 4, 85748 Garching, Germany}

\author{Michelle Lienhart}
\affiliation{Walter Schottky Institut, School of Natural Sciences, and MCQST, Technische Universität München, Am Coulombwall 4, 85748 Garching, Germany}

\author{Christopher Thalacker}
\affiliation{Walter Schottky Institut, School of Natural Sciences, and MCQST, Technische Universität München, Am Coulombwall 4, 85748 Garching, Germany}

\author{Jonathan Finley}
\affiliation{Walter Schottky Institut, School of Natural Sciences, and MCQST, Technische Universität München, Am Coulombwall 4, 85748 Garching, Germany}

\author{Matthias Florian}
\affiliation{Department of Electrical Engineering and Computer Science, University of Michigan, Ann Arbor, Michigan 48109, USA}

\author{Christopher Gies}
\affiliation{Institute for Theoretical Physics and Bremen Center for Computational Material Science, University of Bremen, Bremen, Germany}

\date{\today}

\begin{abstract}
Electrically controllable quantum-dot molecules (QDMs) are a promising platform for deterministic entanglement generation and, as such, a resource for quantum-repeater networks. We develop a microscopic open-quantum-systems approach based on a time-dependent Bloch-Redfield equation to model the generation of entangled spin states with high fidelity. The state preparation is a crucial step in a protocol for deterministic entangled-photon-pair generation that we propose for quantum repeater applications. Our theory takes into account the quantum-dot molecules' electronic properties that are controlled by time-dependent electric fields as well as dissipation due to electron-phonon interaction. We quantify the transition between adiabatic and non-adiabatic regimes, which provides insights into the dynamics of adiabatic control of QDM charge states in the presence of dissipative processes. From this, we infer the maximum speed of entangled-state preparation under different experimental conditions, which serves as a first step towards simulation of attainable entangled photon-pair generation rates. The developed formalism opens the possibility for device-realistic descriptions of repeater protocol implementations.
\end{abstract}

\maketitle

\section{Introduction}\label{sec:intro}
In earth-bound quantum-network implementations, losses in optical fibres limit the maximum distance of quantum-information transfer to a few hundred kilometers \citep{yin2016measurement, boaron2018secure}. For applications such as quantum key distribution, fibre-channel loss is the limiting factor for secret-key generation rates. Quantum repeaters are seen as an enabling technology to extend the repeaterless limit in principle indefinitely by daisy-chaining repeater cells \cite{briegel1998quantum, van2020extending}. Within each cell, entanglement swapping is used to transfer incoming quantum information onto a new set of entangled qubits. Currently, different technological concepts are being considered as hardware platforms for repeater cells \cite{van2020extending, sangouard2011quantum, behera2019demonstration}. Self-assembled semiconductor quantum-dot molecules are among them, with the possibility to host entangled singlet-triplet spin states that are hybridized between the quantum dots (QDs), offering excellent single-photon emission and photon extraction efficiencies \citep{ hennessy2007quantum, jennings2020self}. The orbital hybridization of the vertically stacked quantum dots forming the molecule can be tuned via an electrical field, allowing the precise tuning of the electronic states on a picosecond timescale \citep{krenner2005direct}. Importantly, a sweet spot exists, in which the electronic states are protected against external-field noise. Operating the QDM in this regime leads to coherence times that are up to two orders of magnitude larger than those of single-spin qubits \citep{weiss2012coherent, hiltunen2015charge, tran2022enhanced}.
QDMs are also considered as a suitable platform for the generation of repeater graph states \citep{economou2010optically, buterakos2017deterministic} for all-optical memoryless quantum-repeater implementations \cite{azuma2015all, li2019experimental}.

In this work, we propose a protocol for the deterministic generation of entangled photon-pairs using QDMs and introduce an open-quantum-system approach to simulate protocol sequences on a microscopic level. We explicitly account for the coupled dynamics of the QDM electronic-state excitations, dissipative influences due to electron-phonon interaction, and time-dependent electric fields to exert external control by adapting a Bloch-Redfield equation to open quantum systems that are weakly coupled to the environment \citep{breuer2002theory, delgado2017spin}. We extend the formalism to account for an explicit time dependence in the Hamiltonian and, thereby, dynamically changing eigenstates of the QDM system. The Bloch-Redfield equation is integrated to obtain the system's response in presence of a changing electric field, allowing to quantify the crossover from adiabatic to non-adiabatic dynamics under the influence of dissipation as described by the adiabatic theorem \citep{born1928beweis, kato1950adiabatic}.
In the adiabatic regime, where the switching happens slow enough for the system to stay in its eigenstate, we find that the system's response to the changing electric field can be used to generate entangled states with high fidelity, whereas the onset of non-adiabatic behavior prohibits controlled state preparation. Our model takes the explicit single-particle wave functions of the QDM into account, allowing us to relate QDM fabrication parameters, such as size, geometry, and tunnel coupling strength, and the electric-field control parameters to attainable switching speeds and fidelities that are of direct relevance for quantum-repeater hardware development.
\begin{figure*}[bhtp]
	\centering
	\includegraphics[width=2\columnwidth]{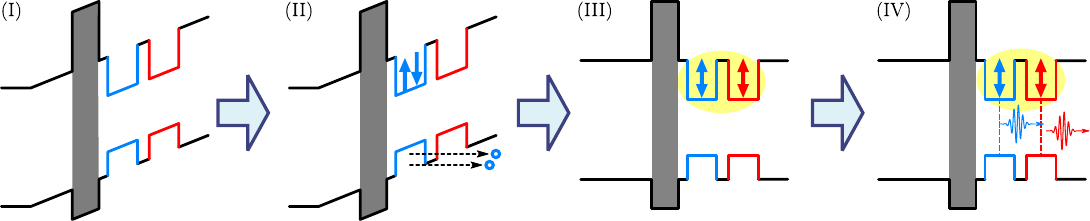}
	\caption{Sketch of the four phases (I)-(IV) of the proposed quantum repeater protocol based on QDMs. (I) The QDM is initialized in a zero-charge state with a large electric-field bias. (II) Deterministic excitation of a two-electron state by external laser pulses and a tunneling ionization process of electron-hole pairs. (III) Redistribution of the charges by adiabatic control, creating delocalized entangled spin states. (IV) Entangled-photon pair generation by scattering of the photons off the QDM charge state.}
	\label{fig:protocol}
\end{figure*}
We begin by outlining the proposed quantum repeater protocol for the generation of entangled photon pairs using QDMs in the following section. Also, the microscopic description of single and two-electron QDM charge states is outlined in Sec.~\ref{sec:qd_qit}.
In Sec.~\ref{sec:redfield} we introduce the Bloch-Redfield formalism and extend it to account for time-dependent Hamiltonians controlled by external electric fields and the time-varying influence of the electron-phonon interaction. In Sec.~\ref{sec:one_carrier}, the formalism is then applied to investigate switching sequences for a single-electron charge state. We use this case to illustrate the main physics and relate the observed diabatic and adiabatic system dynamics to the Landau-Zener model \citep{zener1932non}. 

QDM-based quantum repeater protocols \citep{calderon2019fast, economou2012scalable} rely on the preparation of a delocalized two-electron singlet state in a sweet spot by adiabatic switching using an electric field. This case is considered in Sec.~\ref{sec:two_carriers}, providing an important step towards quantum-mechanical simulations of full gate sequences for quantum technologies in general and quantum repeaters in particular.

\section{Quantum dot molecules for quantum information technology}\label{sec:qd_qit}

Semiconductor QDMs consist of two vertically stacked semiconductor QDs separated by a tunneling barrier a few nm thick. This barrier allows charge carriers to tunnel between the two dots, leading to the formation of delocalized, molecule-like states \citep{doty2008optical}. We focus on InAs QDs, although our formalism can be applied to a wide variety of material systems and platforms.
The energy spectrum and electronic properties of the QDM can be tuned by an external, time-dependent electric field $F(t)$\citep{krenner2005direct, skold2013electrical}, which is realized by embedding the QDM into a p-i-n diode \citep{greilich2011optical}.

We propose a protocol that uses QDMs  for the generation of entangled photon pairs while preserving the possibility to tune the tunnel coupling and the charge states individually. It consists of a four-phase electrical and optical sequence. Figure~\ref{fig:protocol} shows the charge configuration and band alignment of a QDM for the four phases of the protocol. During phase (I), the QDM is initialized in a zero-charge state. A high electric field is applied, causing all carriers to tunnel out of the QDM \citep{bopp2022quantum}. In phase (II) the QDM is deterministically charged with two electrons. Two successive laser pulses create two electron-hole pairs in the lower QD. The interplay between the applied electric field and a higher bandgap tunneling barrier (gray area) leads to a tunneling ionization process of the quantum dot. The two electrons are prepared in a spin-singlet state. Phase (III) redistributes the two charges to both QDs by adiabatic switching of the electric field. Finally, spin-photon entanglement is performed in phase (IV) following established protocols for single QDs \citep{de2012quantum, gao2012observation}.

\begin{figure}[htbp]
	\centering
	\includegraphics[width=\columnwidth]{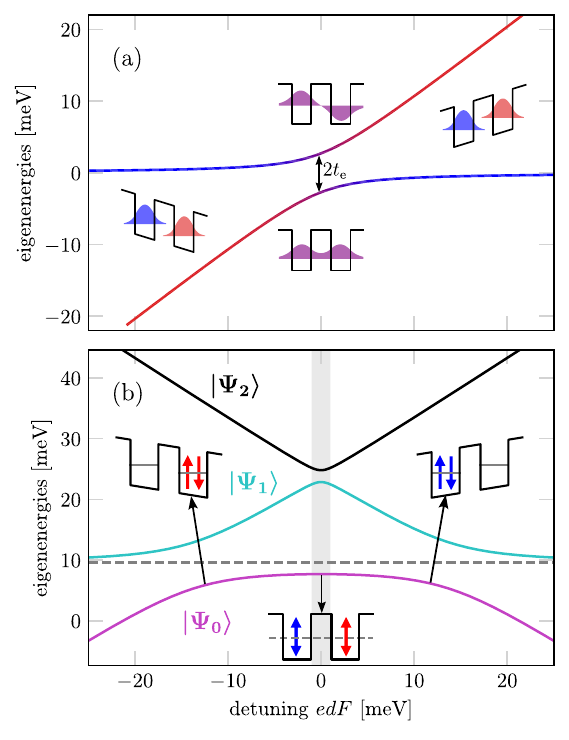}
	\caption{Energy spectrum of a QDM charged with one electron (a) and two electrons (b) as a function of the applied external electric field. In the one-electron case, the admixtures of the localized basis states are color coded, blue representing $\ket{\psi_\mathrm{B}}$ and red representing $\ket{\psi_\mathrm{T}}$. For the two-electron case, the dashed line represents the triplet energy $V_\mathrm{BT}$.}
	\label{fig:electron_energies}
\end{figure}

With the theory developed in this work we study the electron dynamics induced by the electrical switching during the redistribution phase (III). For the description of the QDM dynamics we use a basis of single-electron and two-electron charge states, with no holes residing in the dots. Similarly, hole spins can be employed in this context, and the initialization of such states has recently been demonstrated \citep{scheibner2007spin, bopp2022quantum}.
\\
For the microscopic description of electrons in the QDM, we employ an effective model using basis states $\ket{\psi_\mathrm{B}}$ and $\ket{\psi_\mathrm{T}}$ of the electron residing in the bottom (B) or top (T) dot, respectively. While more sophisticated methods like $\mathbf{k}\cdot\mathbf{p}$ calculations are possible, the above effective model has been used successfully to describe experimental spectra in QDMs \citep{schall2021bright}. For a single electron in the QDM, the Hamiltonian reads
\begin{equation}
H_\mathrm{1e} = \begin{pmatrix}
0 & t_\mathrm{e} \\
t_\mathrm{e} & edF(t)
\end{pmatrix}
\end{equation}
where $edF(t)$ is the energy offset between top and bottom dot due to the electric field $F(t)$ with the electron charge $e$ and inter-dot separation $d$. The tunneling matrix element $t_\mathrm{e}$ between the two dots can be calculated from the single-particle wave functions. We approximate these wave functions by solving an effective Hamiltonian in envelope-function approximation numerically, for more details see Appendix \ref{app:wave_functions}. Without loss of generality, we chose the energy of the bottom dot to be zero at all field strengths. If a strong bias is applied, the electronic eigenstates of the QDM are approximately equal to the localized basis states. At an electric field $edF = 0$, the eigenstates become delocalized, resulting in the molecular states $\ket{\Psi_0} = (\ket{\psi_\mathrm{B}}+ \ket{\psi_\mathrm{T}})/\sqrt{2}$ and $\ket{\Psi_1} = (\ket{\psi_\mathrm{B}}- \ket{\psi_\mathrm{T}})/\sqrt{2}$. Figure~\ref{fig:electron_energies}(a) shows single-electron energies as a function of electric field tuning with a schematic visualization of the eigenfunctions. The tunnel coupling leads to an avoided crossing with energy separation $2 t_\mathrm{e}$. In the following we denote the eigenstates of the underlying Hamiltonian as $\ket{\Psi_i}$, where $\ket{\Psi_0}$ always indicates the lowest-energy eigenstate.

In the case of a QDM charged with two electrons, we use the basis $\ket{xy;S}$, where $x,y = \{ B,T \}$ denote the location of each electron as being in the bottom or top QD, and $S=\{ s,$ $\tau$\,=\,$0$, $\tau$\,=\,$\pm1 \}$ denotes the electrons' spin state, $s$ for the singlet and $\tau$ for the triplet states. The parity of the spatial wave function is given by the opposite of the spin wave functions, since the total wave function has to be antisymmetric under particle exchange. We restrict our description to the singlet subspace $\ket{BB;s}$, $\ket{BT;s}$ and $\ket{TT;s}$. We neglect the corresponding degenerate triplet states $\ket{BT;0}$, $\ket{BT;+}$ and $\ket{BT;-}$ in our description, as they are coupled to the singlet states only by exchange interaction in the presence of holes, or nuclear as well as external magnetic fields, which are small \citep{takagahara1993effects, skold2013electrical, chekhovich2013nuclear}. Given this basis, the Hamiltonian reads
\begin{align}
H_\mathrm{2e}
 = \begin{pmatrix}
V_\mathrm{BB} -edF & -\sqrt{2}t_\mathrm{e} & 0 \\
-\sqrt{2}t_\mathrm{e} & V_\mathrm{BT} & -\sqrt{2}t_\mathrm{e} \\
0 & -\sqrt{2}t_\mathrm{e} & V_\mathrm{TT} +edF \\
\end{pmatrix},
\end{align}
where $V_{ij}$ are the (diagonal) Coulomb matrix elements. More details and derivation of the Coulomb matrix elements are given in Appendix \ref{app:cme}.

Figure~\ref{fig:electron_energies}(b) shows the two-particle energy spectrum. Colored, solid lines show the singlet states, while the dashed line indicates the degenerate triplet states which are insensitive to the electric field and do not couple to other states. The spin triplets are energetically separated from the delocalized singlet state due to exchange interaction among the two electrons. At $edF = 0$, the exchange interaction between the singlet and triplet states is minimal. This is referred to as a \textit{sweet spot}, where the singlet-triplet subspace is protected from electric field fluctuations to first order, as the derivative of the singlet-triplet energy splitting with respect to the field variation vanishes at this point (cf. Figure~\ref{fig:electron_energies}). Preparing singlet or triplet states at the sweet spot extends their coherence time by over an order of magnitude up to hundreds of ns \citep{ weiss2012coherent, tran2022enhanced}.

\section{Bloch-Redfield-Formalism and Electron-Phonon-Interaction}\label{sec:redfield}
The interaction of electrons to the surrounding phonons cause effects like phonon-assisted tunneling and
interdot relaxation \citep{lopez2005phonon, daniels2013excitons}. To describe these effects, we first introduce the well-established Bloch-Redfield equation used to describe open quantum system. We then extend it to a general, explicitly time-dependent Hamiltonian and apply it to our system of confined electrons in a QDM surrounded by a thermal phonon bath.

The general formulation of the Bloch-Redfield equation is given for a Hamiltonian of the form $H = H_\mathrm{S} + H_\mathrm{B} + H_\mathrm{I}$, where $H_\mathrm{S}$ and $H_\mathrm{B}$ only act on the system and the environment, respectively. The coupling between them is assumed to be weak and described by the interaction Hamiltonian
\begin{equation}\label{eq:Hi_general}
H_\mathrm{I} = \sum_{\alpha} A_\alpha \otimes B_\alpha,
\end{equation}
where the operators $A_\alpha$ only act on the system and $B_\alpha$ only act on the environment. The time evolution of the reduced system density matrix is then given by the Bloch-Redfield equation \citep{breuer2002theory}
\begin{align}
\frac{\d}{\d t} \rho_\mathrm{S} = &-\frac{\i}{\hbar} [H_\mathrm{S}, \rho_\mathrm{S}] \nonumber \\ 
&+ \sum_{\alpha\beta, \omega_{ij}} \gamma_{\alpha\beta} (\omega_{ij}) \mathcal{D}[A_\alpha(\omega_{ij}), A_\beta(\omega_{ij})]\rho_\mathrm{S}.\label{eq:BR_equation}
\end{align}
The  first term describes the unitary time evolution induced by the system Hamiltonian $H_\mathrm{S}$, while the dissipator $ \mathcal{D}[X,Y]\rho_\mathrm{S} \equiv (Y \rho_\mathrm{S} X^\dagger - X^\dagger Y \rho_\mathrm{S} + \mathrm{h.c.})/2 $ includes dissipative effects caused by the coupling to the environment. The sum runs over all transition frequencies $\omega_{ij} = (E_i - E_j)/\hbar$, where the indices $i,j$ enumerate the system Hamiltonian's eigenenergies. For a given transition frequency $\omega_{ij}$, $A_\alpha(\omega_{ij}) = \ket{\Psi_j}\bra{\Psi_j}A_\alpha\ket{\Psi_i}\bra{\Psi_i}$ is the corresponding eigenoperator, describing a transition from the system eigenstate $\ket{\Psi_i}$ to $\ket{\Psi_j}$. $\gamma_{\alpha\beta}(\omega) = \hbar^{-2}\int_{-\infty}^\infty \d \tau\, \e^{\i\omega \tau} \braket{B_\alpha^\dagger(\tau) B_\beta(0) }$ is the spectral function, which is evaluated at the transition frequencies. This is valid under Born-Markov approximation, where one assumes a weak system-bath interaction and that the bath correlations decay sufficiently fast, allowing a coarse graining in time \citep{breuer2002theory, haug2008quantum}.

The above formulation is valid for a time-independent system, where the system Hamiltonian does not explicitly depend on time. In order to simulate quantum gate sequences, the electric field and, therefore, the system Hamiltonian and its eigenstates vary over time, in which case Equation~\eqref{eq:BR_equation} is not applicable anymore. Although the interaction Hamiltonian stays intrinsically time independent, it is often convenient to represent $H_\mathrm{I}$ in the eigenbasis of $H_\mathrm{S}(t)$. In this case,  the transition operators $A_\alpha(\omega_{ij})$ as well as the spectral functions $\gamma_{\alpha\beta}(\omega)$  become explicitly time-dependent, and the spectral function needs to be replaced by its time-dependent counterpart, i.e.
\begin{equation}\label{eq:replacement}
\gamma_{\alpha\beta}(\omega, t)= \hbar^{-2}\int_{-\infty}^\infty \d \tau\, \e^{\i \omega \tau} \braket{B_\alpha^\dag(t,\tau)B_\beta(t, 0)}.
\end{equation}
$B_\alpha(t, \tau) = \e^{\i H _\mathrm{B} \tau/\hbar}B_\alpha(t,0)\e^{-\i H _\mathrm{B} \tau /\hbar }$ depends on two times. The dependence on $\tau$ is caused by the time evolution in the interaction picture, while $t$ indicates the time at which the operator is projected onto the time-dependent eigenbasis of $H_\mathrm{S}$. Equation~\eqref{eq:replacement} is valid assuming $H_\mathrm{S}(t)$, and with it the eigenoperators $A_\alpha(\omega_{ij})$, only vary slowly during the environments correlation time,~i.e. the time in which the bath correlation functions $C_{\alpha\beta}(\tau, t) = \hbar^{-2}\braket{B_\alpha^\dag(t,\tau)B_\beta(t,0)}$ decay. A more detailed discussion of this assumption and its validity regarding our system can be found in the Appendix. Since the spectral function is evaluated at the transition frequencies $\omega_{ij}$, the time dependence of the system's energies introduce another temporal dependence on the transition rates. 
While the Bloch-Redfield equation is of Lindblad type due to the secular approximation \citep{breuer2002theory}, the equation we derived differs from the standard Lindblad form, since both the transition rates $\gamma_{\alpha\beta}(\omega, t)$ and the jump operators $A_{\alpha}(\omega, t)$ are derived microscopically and change in time.

To include the effects of electron-phonon scattering, we start with the general form of the interaction Hamiltonian \citep{gawarecki2010phonon}
\begin{equation}\label{eq:Hint}
H_\mathrm{e-ph} = \sum_{nm} a_{n}^\dag a_{m}\sum_{s,\textbf{q}}F_{s,nm}(\textbf{q})(b_{s,\textbf{q}}+ b_{s,-\textbf{q}}^\dag),
\end{equation}
where $a_{n}^{(\dagger)}$ is the annihilation (creation) operator for an electron in the single-particle state $\ket{\psi_n}$, $b_{s,\textbf{q}}^{(\dagger)}$ are annihilation (creation) operators for phonons with wave vector $\textbf{q}$ and phonon branch $s$, and $F_{s,nm}(\textbf{q})$ is the corresponding interaction matrix element.
It is convenient to project the interaction Hamiltonian onto the eigenbasis $\{ \ket{\Psi_i} \}$ of $H_\mathrm{S}$:
\begin{equation}
H_\mathrm{e-ph} = \sum_{ij} \ket{\Psi_i}\bra{\Psi_j}\sum_{s,\textbf{q}}G_{s,ij}(\textbf{q})(b_{s,\mathbf{q}}+ b_{s,-\mathbf{q}}^\dag),\label{eq:H_ep_projection}
\end{equation}
where
\begin{equation}
G_{s,ij}(\mathbf{q}) = \sum_{nm} \bra{\Psi_i} a_{n}^\dag a_{m} \ket{\Psi_j} F_{s,nm}(\mathbf{q})
\end{equation}
are the transformed interaction matrix elements in this new basis. This is the required form for Eq.~\eqref{eq:Hi_general}, where we identify $\ket{\Psi_i}\bra{\Psi_j} = A_\alpha$ and $\sum_{s,\textbf{q}}G_{s,ij}(\textbf{q})(b_{s,\mathbf{q}}+ b_{s,-\mathbf{q}}^\dag) = B_\alpha$ with $\alpha = (i,j)$. It now becomes apparent that in this form, the eigenoperators $A_{(i,j)}(\omega_{j^\prime i^\prime}) = \ket{\Psi_{j^\prime}}\braket{\Psi_{j^\prime}| \Psi_j} \braket{\Psi_i | \Psi_{i^\prime}}\bra{\Psi_{i^\prime}} = \ket{\Psi_j}\bra{\Psi_i}\delta_{i i^\prime} \delta_{j j^\prime}$, which simplifies the Bloch-Redfield equation to
\begin{align}
\frac{\d}{\d t} \rho_\mathrm{S} &= -\frac{\i}{\hbar} [H_\mathrm{S}(t), \rho_\mathrm{S}] \nonumber \\
&+ \sum_{\alpha \beta} \gamma_{\alpha \beta} (\omega_\alpha(t), t) \mathcal{D}[A_{\alpha}(t), A_{\beta}(t)]\rho_\mathrm{S}
\end{align}
now carrying an explicit time dependence for the jump operators, the spectral function, and for the transition frequency it is evaluated at. Changing the external electric field changes the way the electrons are able to scatter off phonons as is now reflected in the time-dependent spectral function. If a strong field is applied, the electron energy separation becomes large compared to the energy of the phonons while their wave-function overlap becomes small, both hindering efficient electron-phonon scattering. The spectral function for the phonon bath is given by \cite{may2023charge}
\begin{align}
\gamma_{\alpha\beta}(\omega) = 2\pi\left[ J_{\alpha\beta}(-\omega) n(-\omega) + J_{\alpha\beta}(\omega) (n(\omega) + 1) \right],
\end{align}
where $n(\omega) = {1}/({\e^{\hbar\omega/(k_\mathrm{B}T)} - 1})$ is the Bose-Einstein distribution for the phonons in thermal equilibrium at temperature $T$, and
\begin{equation}
J_{\alpha\beta}(\omega) = \sum_{\mathbf{q},s} G_{s,\alpha}^\ast(\mathbf{q})G_{s,\beta}(\mathbf{q}) \delta(\omega-\omega_{\mathbf{q},s})
\end{equation}
the spectral density with the indices $\alpha$ and $\beta$ enumerating the transitions between eigenstates.
\\
\begin{figure}[htbp]
	\centering

	\includegraphics[width=0.98\columnwidth]{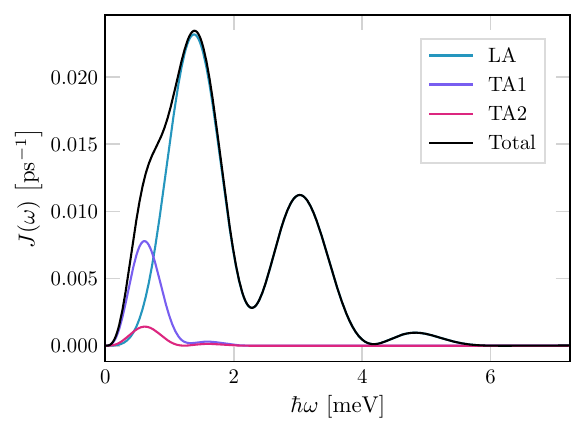}
	\caption{Spectral density for the transition between single electron eigenstates at $F=0$, e.g.~$\alpha = \beta = (0,1)$ broken down into its individual contributions from the different acoustic phonon branches, i.e. the longitudinal branch (LA) and the two transversal branches (TA1 and TA2).}
	\label{fig:spectral_density}
\end{figure}
For the calculations of the electron-phonon interaction matrix elements and the spectral densities, we consider a deformation potential (DP) coupling to the longitudinal acoustic (LA) phonon branch and a piezoelectric potential (PE) to both the LA and the transversal acoustic (TA) branches \citep{gawarecki2010phonon}. Due to the small energy splitting, the optical phonons do not contribute to dissipative effects. The explicit form of the matrix elements and details about the calculation of the spectral density can be found in Appendix~\ref{app:phonon}.\\
Figure~\ref{fig:spectral_density} shows the spectral density and the contributions from the individual phonon branches for the transition between single electron eigenstates. While the largest contribution comes from the deformation potential coupling, the PE coupling to the transversal branches actually dominates for small energies because of the slower speed of sound compared to the longitudinal branch \citep{gawarecki2010phonon}. This contribution has a significant effect on scattering at low detunings for smaller tunnel couplings, which is why it is important to account for both the DP and PE coupling simultaneously.

\section{Single Electron Dynamics} \label{sec:one_carrier}
We now apply the introduced framework to investigate the transition from adiabatic to non-adiabatic dynamics of the system in dependence of the electric-field switching speed and the tunnel coupling. The adiabtic theorem \cite{born1928beweis, kato1950adiabatic} states that under an external time-dependent perturbation, a quantum mechanical system will approximately stay in its instantaneous eigenstate as long as the change occurs slowly compared to the energy splitting between the corresponding eigenenergy and the remaining eigenspectrum. Our theory allows to quantify how dissipative processes modify the dynamics of the system predicted by the adiabatic theorem. 
\\
The time dependence of the electric field varying smoothly from $f_\mathrm{max}$ to $-f_\mathrm{max}$ is modeled as
\begin{equation}\label{eq:switching_tanh}
F(t) = -f_\mathrm{max} \tanh(kt),
\end{equation}
where $f_\mathrm{max}$ is the maximum absolute value of the electric field at $t=\pm \infty$ and $k$ is a parameter determining the speed of the electric-field variation.
\\In the following analysis, we refer to $v = k d_\mathrm{i} f_\mathrm{max} $ as the switching speed, with $d_\mathrm{i}= \SI{200}{nm}$ as the size of the diodes intrinsic region \citep{bopp2022quantum}. We note that the switching speed can be scaled by an arbitrary factor to account for differently sized diodes.
\\
To illustrate the interplay of various effects and mechanisms, we simulate the following switching sequence:
\begin{enumerate}
\item A large initial electric field $f_\mathrm{max}$ is used, resulting in a large energy offset and eigenstates that are localized in the separate QDs (see Sec. \ref{sec:qd_qit}), such that $\ket{\Psi_0} \approx \ket{\psi_\mathrm{B}}$ and $\ket{\Psi_1} \approx \ket{\psi_\mathrm{T}}$. The higher-energy eigenstate $\ket{\Psi_1}$ is used as initial state.
\item The electric field is varied according to Eq.~\eqref{eq:switching_tanh} inverting its orientation, such that $\ket{\Psi_0} \approx \ket{\psi_\mathrm{T}}$ and $\ket{\Psi_1} \approx \ket{\psi_\mathrm{B}}$ (cf.~Figure~\ref{fig:electron_energies}).
\end{enumerate}
\begin{figure}[htbp]
	\centering
	\includegraphics[width=1\columnwidth]{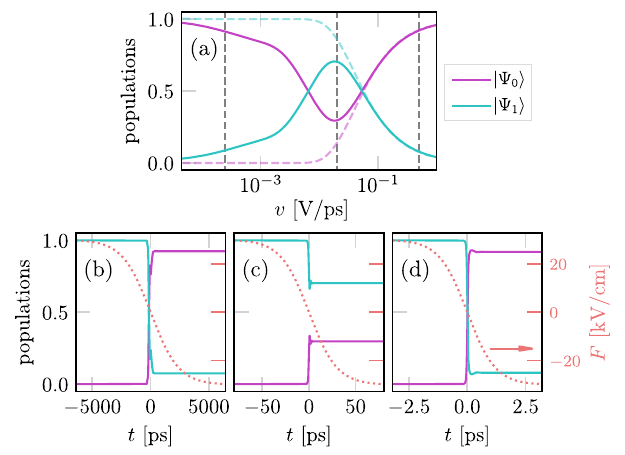}
	\caption{(a) Resulting eigenstate population as a function of the switching speed after the switching of the electric field is completed. Solid and dashed lines compare the case with and without dissipation, respectively. A fixed $t_\mathrm{e} = \SI{0.5}{meV}$ and $T=\SI{10}{K}$ is used. (b)-(d) Time evolution of the occupations for slow ($v=2.5\cdot10^{-4}\, {\mathrm{V}}/{\mathrm{ps}}$), intermediate ($v=\SI{0.02}{{V}/{ps}}$) and fast ($v=\SI{.5}{{V}/{ps}}$) switching of the electric field, which is shown as dotted lines. Note the varying scale of the $t$-axis. The vertical dashed lines in (a) indicate the three switching speeds shown in (b)-(d).} 
	\label{fig:1e_figure}
\end{figure}
In the following, we study the system's response to the temporal change of the eigenstates as illustrated in Figure~\ref{fig:electron_energies}(a).
Figure~\ref{fig:1e_figure}(a) shows the population of the eigenstates $\ket{\Psi_i}$ after switching as a function of the switching speed $v$ (solid lines). The results are compared to the dynamics without dissipation (dashed lines). In the latter case, at low switching speeds the systems remains in its initial eigenstate $\ket{\Psi_1}$ as predicted by the adiabatic theorem, i.e.\ the electron is adiabatically transferred from one QD to the other. For faster switching, the system's state experiences a partial transition between eigenstates, as the electron is not able to adapt to fast changes in the electric field.
When dissipation via interaction with phonons is included in the analysis, slow switching  leaves sufficient time for dissipation to take place, resulting in a decay of the higher-energy state $\ket{\Psi_1}$ to the lower one. The two competing effects, dissipation and non-adiabatic dynamics, lead to a maximum final state population of 0.7 for $\ket{\Psi_1}$.

The time evolution of the populations of $\ket{\Psi_0}$ and $\ket{\Psi_1}$ during the switching are shown in panels (b)-(d) of Figure~\ref{fig:1e_figure} for slow, intermediate, and fast switching speeds. The electric field is shown as dotted line. In all three cases, the population transfer happens in a tight window around $F(t) \approx 0$, where the two QDs are in resonance. 
In the case of slow switching (panel (b)), the observed population change is a consequence of the interaction between the electron and the phonon bath at $T=10\,\mathrm{K}$. For large detuning of the QDs, the necessary transition energy is too large and there are no phonon states to scatter into. As a consequence, electron-phonon scattering primarily occurs around the anticrossing region, where the transition energy of the electron is of the order of typical phonon energies, as can be inferred from the phonon-spectral function in Figure~\ref{fig:spectral_density}.

For fast switching (panel (d)), although the outcome appears similar, the transfer of populations between eigenstates occurs due to non-adiabatic transitions caused by the rapid switching process. The electron remains in the top QD throughout the process, corresponding to the higher-energy state before and the lower-energy state after the switching. In this case, dissipative effects are negligible, since the duration of the entire switching process is shorter than the timescales associated with electron-phonon interactions. We note that, although no dissipative effects are predicted from the Bloch-Redfield equation at switching speeds ${v \gtrsim \SI{0.1}{V/ps}}$, non-Markovian effects taking place on ultra-short timescales that are not included in our model may play an additional role \citep{cosacchi2018path}.

The two limiting cases of slow and fast switching illustrate the role of the two competing processes that determine the final state population: thermalization due to phonon scattering and non-adiabatic dynamics. The behavior at  intermediate switching speeds (panel (c)) is determined by a partial contribution of both, resulting in a mixture of eigenstates.

\begin{figure}[htbp]
	\centering
	\includegraphics[width=0.98\columnwidth]{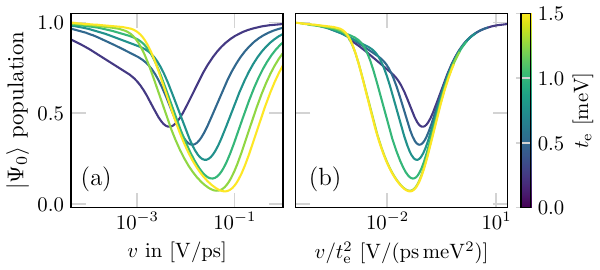}
	\caption{(a) Eigenstate population for $\ket{\Psi_0}$ after switching as a function of the switching speed $v$ and for different tunneling strengths $t_\mathrm{e}$ (color bar). In (b), the axis has been rescaled as a function of $v/t_\mathrm{e}^2$ , allowing for comparison with the Landau-Zehner model.}
	\label{fig:one_e_v_over_t}
\end{figure}

For a better understanding of the role of the tunneling strength in the switching process, Figure~\ref{fig:one_e_v_over_t}(a) shows the final population of the target state $\ket{\Psi_0}$ for varying $t_\mathrm{e}$. It can be observed that increasing the value of $t_\mathrm{e}$ shifts the transition to the non-adiabatic regime towards higher switching speeds.
The behavior can be understood when comparing the results to the Landau-Zener (LZ) model \citep{zener1932non, landau1932theorie}. Neglecting dissipative effects, it predicts that in a coupled two-level system whose energy levels are shifted to each other linearly in time, the probability for a non-adiabatic transition scales as $v/t_\mathrm{e}^2$. 
In panel (b), we show the results of panel (a) on an axis that has been rescaled by $t_\mathrm{e}^2$. In this representation, it becomes apparent that the population behavior indeed follows the expected scaling with $v/t_\mathrm{e}^2$ at fast switching in agreement  with the LZ model. This suggests that a comparison between our system and the LZ model is justified in this regime, where dissipation has negligible effects. The agreement breaks down for slower switching speeds. This is expected, as here, we have already identified dissipation to be the main cause for the transition. 

To conclude this section, our formalism allows to predict the system dynamics under time-dependent switching of the eigenstates across different regimes and timescales and on the basis of material specific properties, such as the tunnel coupling strength and the explicit form of the QMD wave functions, or even for other material systems. In the fast switching regime, it recovers the physics of the LZ model. At slow switching speeds, the solid foundation of treating electron-phonon interaction enables to obtain results that are dominated by scattering and dissipation. In that sense, our modified Bloch-Redfield approach is well suited for the simulation of quantum protocols, especially at the intersection of different time scales.

\section{Deterministic generation of delocalized singlet states} \label{sec:two_carriers}
In this section, we shift our focus to doubly charged QDMs and the preparation of delocalized, entangled singlet states in the sweet spot as depicted in Figure~\ref{fig:electron_energies}(b). The deterministic generation of such entangled electron states is essential to unlock the potential of QDMs for various quantum-technological applications, including their utilization as S-T qubits \citep{li2000single} and specifically in the entangled-photon generation protocol of we proposed in Sec.~\ref{sec:qd_qit}.

For our calculation we assume preparation of the initial state $\ket{\Psi_0} \approx \ket{BB;s}$ with an electric field offset of $2f_\mathrm{max}$, such that both electrons are localized in the bottom QD. The two QDs are then switched into resonance at $F=0$ with the time dependence of the electric field modeled as $F(t) = f_\mathrm{max} (\tanh(kt) -1)$. Defining $2f_\mathrm{max}$ as the maximum field strength ensures that the switching speed is consistent with the one used in the previous section.

\begin{figure}[htbp]
	\centering
	\includegraphics[width=\columnwidth]{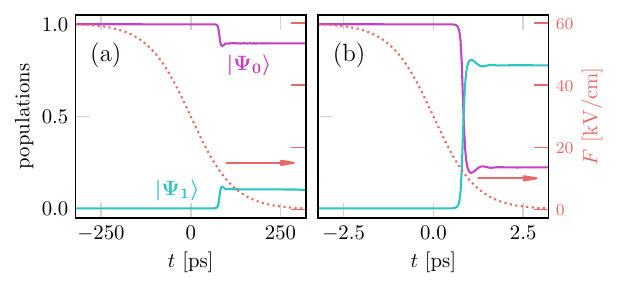}	
	\caption{Time evolution of a system in a two electron charge state during a switching process for (a) slow ($v=\SI{.003}{{V}/{ps}}$) and (b) fast ($v=\SI{0.5}{{V}/{ps}}$) switching speeds. The system is initially prepared in the state $\ket{\Psi_0}$.}
	\label{fig:two_electron_time_evo}
\end{figure}

Figure~\ref{fig:two_electron_time_evo} provides insight into the time evolution of the population in both the slow (a) and fast (b) switching regime for a phonon-bath temperature of $\SI{10}{K}$ and $t_\mathrm{e}=\SI{0.5}{meV}$. In the slow (adiabatic) switching regime, the system predominantly remains in the lowest-energy state $\ket{\Psi_0}$, but experiences thermal mixing caused by electron-phonon scattering into higher-lying states at finite temperatures, reducing the attainable fidelity below one. In the fast switching case, dissipative effects only play a negligible role but the switching happens non-adiabatically, causing a transition from $\ket{\Psi_0}$ to $\ket{\Psi_1}$.

To successfully implement quantum repeaters with QDMs, one aims at maximizing the preparation rate of entangled electron states with high fidelity for subsequent entangled photon-pair generation (cf.~Figure~\ref{fig:protocol}). Achieving this goal requires to determine an optimal switching speed that strikes a balance between maintaining control over the system by staying in the adiabatic regime and achieving maximum repetition rates.

\begin{figure}[htbp]
	\centering
	\includegraphics[width=0.95\columnwidth]{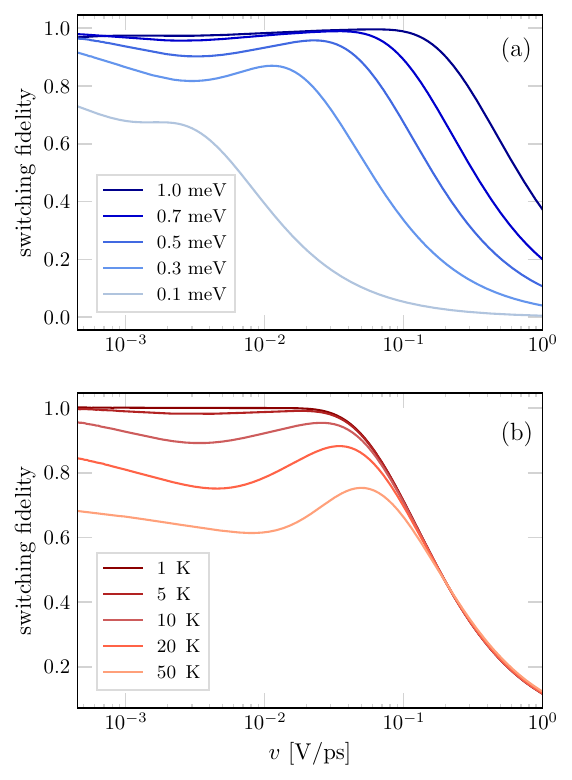}
	\caption{Switching fidelity $\bra{\Psi_0}\rho\ket{\Psi_0}$ (ground state population after switching) as a function of the switching speed $v$ for (a) $T=\SI{10}{K}$, and different tunnel couplings and for (b) different temperatures and a constant tunnel coupling $t_\mathrm{e} = \SI{0.5}{meV}$.}\label{fig:two_electron_switching_speeds}
\end{figure}

In Figure~\ref{fig:two_electron_switching_speeds} we investigate the achievable fidelities for this switching operation, i.e.~the population of the state $\ket{\Psi_0}$ after the switching step, as a function of the electric field switching speed. We do this by varying both the tunnel coupling strength in panel (a) and the temperature in panel (b). 
Without any dissipation, the results are similar to the case of the dissipationless one-electron charge state (cf. the dashed lines in Figure~\ref{fig:1e_figure}\,(a)). The system remains in its eigenstate for slow switching but experiences non-adiabatic transitions for faster switching, leading to a decrease in the final fidelity. This is shown in panel (a) for a large value of the tunneling strength $t_\mathrm{e}=\SI{1.0}{meV}$, where dissipation does not have a significant effect on the fidelity, even for slow switching. The reason for this is that the energy splitting at the anticrossing is proportional to $t_\mathrm{e}$. Therefore, for high tunneling strengths $t_\mathrm{e} \gg k_\mathrm{B}T$ no scattering occurs, as thermal excitations are improbable to overcome the splitting energy. 
For weaker tunnel couplings, we observe two effects: The transition into the non-adiabatic regime shifts towards slower switching speeds, reflecting the $t_\mathrm{e}$-dependence of the transition to non-adiabatic dynamics of the Landau-Zener model as discussed in the context of Sec.~\ref{sec:one_carrier}. At the same time, the fidelity is decreased for smaller $t_\mathrm{e}$ also for the slowest switching speed. This is explained by interactions with thermal phonons, which start to overcome the decreasing energy splitting leading to scattering into higher energy states.

Results in panel (b) are explained in a similar way as temperature is varied for a constant tunneling strength. For fast switching there is only negligible influence of the temperature and fidelities are low due to state population transfer in the non-adiabatic regime. As thermal energies $k_\mathrm{B}T$ become comparable to the energy splitting, electron phonon scattering decreases the fidelities of state preparation in the adiabatic regime.

\begin{figure}[htbp]
	\centering
	\includegraphics[width=0.95\columnwidth]{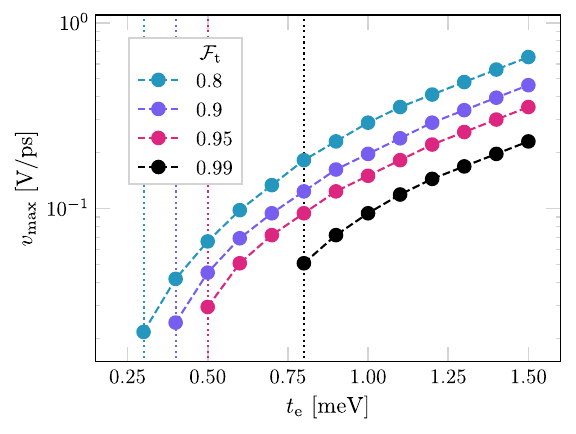}
	\caption{Fastest possible switching speed $v_\mathrm{max}$, for which a target fidelity $\mathcal{F}_\mathrm{t}$ can still be reached. Different curves represent different target fidelities at $T=\SI{10}{K}$.}\label{fig:two_electron_max_speeds}
\end{figure}

In Figure~\ref{fig:two_electron_max_speeds} we show, as a function of the tunneling strength $t_\mathrm{e}$, the fastest possible switching speeds, at which desired target fidelities $\mathcal{F}_\mathrm{t}$ can still be reached. For any fixed fidelity, a stronger tunnel coupling allows for a faster switching while still achieving the target fidelities. We find that depending on the target fidelity, there exist regimes for the tunnel coupling strength, as indicated by the dotted lines, in which these target fidelities are impossible to achieve irrespective of the switching speed. This is due to the increased effect of electron-phonon scattering at slow switching. It is apparent that even high fidelities are achievable for suitable devices, i.e. strongly coupled QDMs. Achieving fidelites of $\SI{99}{\percent}$ requires a tunnel coupling strength of at least $t_\mathrm{e} = \SI{0.8}{meV}$, for which switching as fast as $\SI{0.05}{V/ps}$ is possible.

Our results show that we can efficiently relate device properties such as tunnel coupling strength $t_\mathrm{e}$ and operation temperature to state-preparation fidelities and maximum switching speeds.  Our approach offers a versatile tool for the prediction of fabrication parameters for achieving performance goals of QDMs in quantum repeater applications.

\section{Conclusion}
Quantum-dot molecules are experiencing a revival for their potential in quantum-technology applications. We consider them as a platform for current realizations of quantum repeater protocols. In the protocol implementation that we suggest, time-dependent electric fields are used to switch the QDM energy levels between hybridized and localized eigenstates to deterministically create entangled states of two electrons. The final step of the protocol then sees to the transfer of this spin entanglement to travelling photons.

We calculate attainable rates and fidelities of adiabatic entangled-state preparation, employing a fully quantum-mechanical approach that takes into account the explicit QDM single-particle wave functions and the spectral functions of the electron-phonon interaction. The explicit time dependence of the electric field is captured by solving the Bloch-Redfield equation for time-varying eigenstates and system-bath interaction with the relevant phonon branches. Our results lay down the foundation to further evaluate the capabilities of specific repeater-cell implementations and benchmark them against each other.

For current state-of-the-art realizations of epitaxially grown QDM in switchable electric fields \cite{bopp2022quantum}, we find that entanglement can be created with high fidelity. For future applications, however, switching times will be drastically reduced to tens of ps \citep{mukherjee2020electrically}, in which case we predict that QDM with larger tunnel coupling strength in the range of $t_\mathrm{e} \geq \SI{0.7}{meV}$ will be required to still operate in the adiabatic regime. These results demonstrate the potential of our approach in guiding future experimental developments, and of using QDM in switchable electric fields as a viable platform for quantum-repeater implementations.

The theoretical framework that we provide can be adapted to a variety of different experimental settings and material systems. Different charge states of QDMs can be included in the description as well as external magnetic fields that change the spin dynamics of electrons and holes while also incorporating the effect of electric field noise. While we have evaluated the Bloch-Redfield framework on the foundation of model wave functions, more advanced electronic-state calculations, like $\mathbf{k}\cdot \mathbf{p}$-theory, can be used in a straightforward manner if required.

\section{Acknowledgements}
The groups from Bremen and Munich acknowledge funding from the German Ministry of Science and Education (BMBF) via the project QR.X. M. F. acknowledges support by the Alexander von Humboldt Foundation.

\appendix

\section{Single-particle wave functions and matrix elements}\label{app:wave_functions}
\begin{figure}[htbp]
	\centering
	\includegraphics[width=0.98\columnwidth]{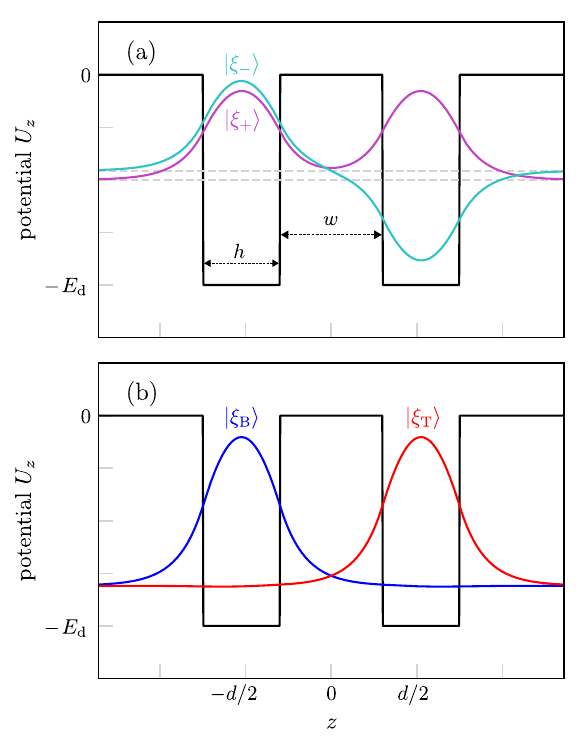}
	\caption{Sketch of the confinement potential in growth direction of the QDM at zero field strength with the depth of the potential well $E_\mathrm{d}$, the tunneling-barrier width $w$ and the height of the QD $h$ with (a) the solution of the effective Schrödinger equation, and (b) the basis functions used to represent the Hamiltonians.}\label{fig:sketch_well}
\end{figure}

In the envelope-function and effective-mass approximation, the envelope wave function $\psi(\mathbf{r})$ obeys the effective Schrödinger equation \citep{haug2009quantum}
\begin{equation}
\left[- \frac{\hbar^2}{2m^\ast}\bigtriangleup + U_\mathrm{conf}(\mathbf{r}) \right] \psi_n (\mathbf{r}) \equiv \mathcal{H} \psi_n (\mathbf{r}) = \varepsilon_n \psi_n (\mathbf{r})
\end{equation}
with the effective mass $m^\ast$ and the confinement potential $U_\mathrm{conf}(\mathbf{r}) = U_\rho(\boldsymbol{\rho}) + U_\mathrm{z}(z)$, which we assume to be separable into an in-plane part $U_\rho(\boldsymbol{\rho})$, and a part in growth direction $U_\mathrm{z}(z)$. We model the in-plane potential to be harmonic $U_\rho(\boldsymbol{\rho}) = {m^\ast \omega_0^2}\rho^2 /2$ and the potential in growth direction $U_z(z)$ as a double finite potential well which we sketch in Figure~\ref{fig:sketch_well}. The basis functions are calculated in the absence of any electric field.

The separability of the potential allows us to separate the wave function into $\psi_n(\mathbf{r}) = \varphi_n(\boldsymbol{\rho})\xi_n(z)$. While we solve for $\xi(z)$ numerically, the in-plane ground state is given by
\begin{equation}
\varphi_0(\boldsymbol{\rho}) = \frac{\beta_\mathrm{e}}{\sqrt{\pi}}\e^{-\frac{\beta_\mathrm{e}^2\rho^2}{2}}
\end{equation}
with the inverse oscillator length $\beta_\mathrm{e} = \sqrt{{\hbar}/({m^\ast \omega_0})}$. The excited states can be neglected, since we are only interested in low temperatures $k_\mathrm{B}T \ll \hbar \omega_0$.
Our basis of choice to represent the Hamiltonian consists of the localized one-dot wave functions $\psi_\mathrm{B/T}(\mathbf{r}) = \varphi_0(\boldsymbol{\rho})\xi_\mathrm{B/T}(z)$. We calculate these basis functions by solving the one-dimensional Schrödinger equation numerically, whose two lowest-energy solutions are the bonding $\ket{\xi_+}$ and antibonding state $\ket{\xi_-}$, which are shown in Figure~\ref{fig:sketch_well}(a).  The localized basis functions shown in Figure~\ref{fig:sketch_well} can then be expressed as a linear combination of those solutions:
\begin{align}
\ket{\xi_\mathrm{B}} &= \frac{1}{\sqrt 2}(\ket{\xi_+} + \ket{\xi_-}) \\
\ket{\xi_\mathrm{T}} &= \frac{1}{\sqrt 2}(\ket{\xi_+} - \ket{\xi_-})
\end{align}
These basis functions are similar to the solutions of a single finite potential well with the important difference that they are orthogonal to each other, making them suitable as a basis for further calculations.

With these basis functions, we can compute the matrix elements for the single-particle Hamiltonian, including an
external electric field that shifts the potential by $ezF$, given by
\begin{equation}
H^{(1e)}_{ij} = \int \d^3 r \, \psi_i^\ast(\mathbf{r}) \left[- \frac{\hbar^2}{2m^\ast}\bigtriangleup + U_\mathrm{conf} (\mathbf{r}) + ezF \right] \psi_j(\mathbf{r}).
\end{equation}
Due to the symmetry of the wave functions, the term $ezF$ gives no contribution to the off-diagonal elements, which simplify to
\begin{equation}
H^{(1e)}_{BT} = \int \d z\, {\xi}_\mathrm{B}^\ast (z) \left[- \frac{\hbar^2}{2m^\ast}\frac{\partial^2}{\partial z^2} + U_z(z) \right] {\xi}_\mathrm{T} (z) \equiv t_\mathrm{e}.
\end{equation}
In good approximation, assuming the basis functions are localized to one dot only (i.e. close to the single-potential-well solutions), the matrix elements only differ by
\begin{equation}
H^{(1e)}_{TT} - H^{(1e)}_{BB} = +edF,
\end{equation}
which can be seen by considering the symmetry of the wave functions. We choose $H^{(1e)}_{BB} = 0$ as the origin of energy w.l.o.g. as mentioned in the main text and, accordingly, $H^{(1e)}_{TT} = edF$. 

\section{Coulomb Matrix Elements}\label{app:cme}
The two-electron singlet subspace is spanned by the following states \citep{scheibner2007spin}:
\begin{align}
\ket{B B; s} &= \ket{\psi_\mathrm{B}}\ket{\psi_\mathrm{B}}(\ket{\uparrow}\ket{\downarrow} - \ket{\downarrow} \ket{\uparrow})/\sqrt{2} \nonumber
\\
\ket{B T; s} &= (\ket{\psi_\mathrm{B}}\ket{\psi_\mathrm{T}} + \ket{\psi_\mathrm{T}}\ket{\psi_\mathrm{B}} ) (\ket{\uparrow}\ket{\downarrow} - \ket{\downarrow} \ket{\uparrow})/2 \nonumber
\\
\ket{T T; s} &= \ket{\psi_\mathrm{T}}\ket{\psi_\mathrm{T}}(\ket{\uparrow}\ket{\downarrow} - \ket{\downarrow} \ket{\uparrow})/\sqrt{2},
\end{align}
while we neglect the triplet states
\begin{align}
\ket{B T; 0} &= (\ket{\psi_\mathrm{B}}\ket{\psi_\mathrm{T}} - \ket{\psi_\mathrm{T}}\ket{\psi_\mathrm{B}} ) (\ket{\uparrow}\ket{\downarrow} + \ket{\downarrow} \ket{\uparrow})/2 \nonumber
\\
\ket{B T; -} &= (\ket{\psi_\mathrm{B}}\ket{\psi_\mathrm{T}} - \ket{\psi_\mathrm{T}}\ket{\psi_\mathrm{B}} ) \ket{\downarrow}\ket{\downarrow} /\sqrt{2} \nonumber
\\
\ket{B T; +} &= (\ket{\psi_\mathrm{B}}\ket{\psi_\mathrm{T}} - \ket{\psi_\mathrm{T}}\ket{\psi_\mathrm{B}} ) \ket{\uparrow}\ket{\uparrow} /\sqrt{2} 
\end{align}
that decouple from the singlet subspace due to the conservation of total spin.

\begin{figure}[htbp]
	\centering
	\includegraphics[width=\columnwidth]{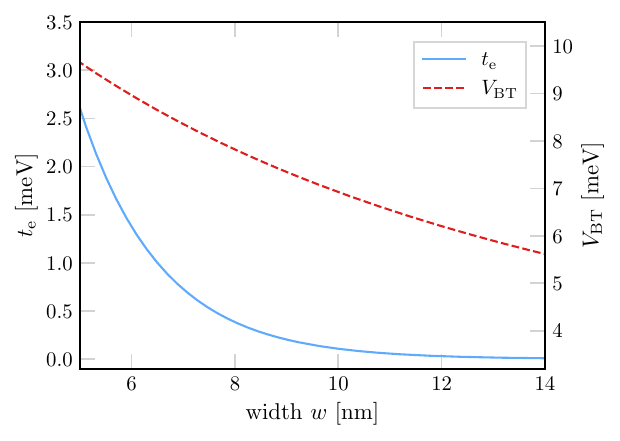}
	\caption{Tunnel matrix element $t_\mathrm{e}$ (blue, solid) and Coulomb matrix element $V_\mathrm{BT}$ (red, dashed) as a function of the barrier width $w$ for a constant quantum dot height $h=\SI{4.5}{nm}$.}
	\label{fig:matrix_elements}
\end{figure}

Since the single-particle matrix elements are known from Sec.~\ref{app:wave_functions}, we now focus on the Coulomb matrix elements (CME). Due to the small overlap between wave functions from different quantum dots, we neglect all CME that feature the overlap of two wave functions from different QDs. The non-negligible CME are $V_\mathrm{TT}$, $V_\mathrm{BB}$ and $V_\mathrm{BT}$, which describe the classical Coulomb repulsion between two electrons. They are given by 
\begin{equation}
V_{ij} = \frac{e^2}{4\pi\epsilon_0\epsilon_\mathrm{r}}\int \d^3r \int \d^3r^\prime |\psi_i^\ast (\mathbf{r})|^2 \frac{1}{|\mathbf{r} - \mathbf{r}^\prime|} |\psi_j^\ast (\mathbf{r}^\prime)|^2
\end{equation}
with $i, j = \{\mathrm{B}, \mathrm{T} \}$ indicating the single-particle eigenstates. The triplet states are eigenstates with energy $V_\mathrm{BT}$, not coupling to other states without external magnetic fields. Doing the same factorization of wave functions $\psi_i(\mathbf{r}) = \varphi_0(\boldsymbol{\rho}) \xi_i(z)$, this equals to
\begin{equation}
V_{ij} = \frac{\pi}{(2\pi)^3} \frac{e^2}{\epsilon_0\epsilon_r} \int \d^2 q_\rho \frac{F_{ij}(q_\rho)}{q_\rho} 
|\bra{\varphi_{0}} \e^{ \i \mathbf{q}_\rho \cdot \boldsymbol{\rho}} \ket{\varphi_{0}}|^2
\end{equation}
in reciprocal space, where
\begin{equation}
F_{ij}(q_\rho) = \int \d z \, |\xi_{i}(z)|^2
\int \d z^\prime \, |\xi_{j}(z^\prime)|^2\e^{-q_\rho |z-z^\prime|}
\end{equation}
is form a factor that accounts for the shape of the wave function in $z$-direction \cite{chow1999semiconductor}. The matrix element of the harmonic ground state is given by
\begin{equation}
|\bra{\varphi_{0}} \e^{ \i \mathbf{q}_\rho \cdot \boldsymbol{\rho}} \ket{\varphi_{0}}|^2 = \e^{-\frac{q_\rho^2}{2\beta_\mathrm{e}^2}}.
\end{equation}
We calculate the form factors numerically using the basis functions introduced in Sec.~\ref{app:wave_functions}. The resulting matrix elements $V_\mathrm{BT}$ and $t_\mathrm{e}$ are shown in Figure~\ref{fig:matrix_elements}. While the tunnel matrix element $t_\mathrm{e}$ decays exponentially with the barrier width due to the exponentially decaying wave functions, $V_\mathrm{BT}$ decays linearly for small QD separation and as $1/w$ for large widths due to the form of the Coulomb interaction.

\section{Electron-Phonon-Interaction}\label{app:phonon}
It is convenient to reduce the full spectral density $J_{\alpha\beta}(\omega)$ to the single-particle spectral densities
\begin{equation}\label{eq:imunu}
I_{\mu\nu}(\omega) = \sum_{\mathbf{q},s} F_{s,\mu}^\ast(\mathbf{q})F_{s,\nu}(\mathbf{q}) \delta(\omega-\omega_{\mathbf{q},s})
\end{equation}
by linearly combining them via
\begin{equation}
J_{\alpha\beta}(\omega) = \sum_{\mu\nu} M_{\alpha\mu}^\ast M_{\beta\nu} I_{\mu\nu}(\omega).
\end{equation}
The Matrix elements that relate these spectral densities are defined as
\begin{equation}
M_{\alpha\mu} =\bra{\Psi_{i_\alpha}} a_{n_\mu}^\dag a_{m_\mu} \ket{\Psi_{j_\alpha}}.
\end{equation}
The indices $\mu = (n_\mu, m_\mu)$ and $\nu = (n_\nu, m_\nu)$ enumerate \emph{single-particle transitions}, i.e.~those described by the operator $a_{n_\mu}^\dag a_{m_\mu}$.
For the calculation of the spectral function defined in Eq.~\eqref{eq:imunu}, the explicit form of the interaction matrix elements is needed, which is \cite{krummheuer2005pure}
\begin{equation}
F_{s,\mu} (\mathbf{q}) = \mathcal{G}_s(\mathbf{q})\mathcal{F}_\mu(\mathbf{q}),
\end{equation}
with the form factor
\begin{equation}
\mathcal{F}_\mu (\mathbf{q}) = \int \d^3 r \, \e^{\i\mathbf{q}\cdot \mathbf{r}} \psi_{n_\mu}^\ast(\mathbf{r})\psi_{m_\mu}(\mathbf{r}).
\end{equation}
The prefactor $\mathcal{G}_s(\mathbf{q})$ is determined by the type of the phonon coupling and  the respective phonon branch \cite{krummheuer2005pure}:
\begin{align}
\mathcal{G}_\mathrm{LA}(\mathbf{q}) =& \sqrt{\frac{\hbar q}{2\rho V c_\mathrm{l}}} D -\i \frac{3}{2}\sqrt{\frac{\hbar}{2 \rho V c_\mathrm{l} q}} \frac{d_\mathrm{p}e}{\epsilon_0\epsilon_\mathrm{r}} \sin(2\vartheta) \sin\vartheta \sin\varphi \nonumber
\\
\mathcal{G}_\mathrm{TA1}(\mathbf{q}) =& -\i \sqrt{\frac{\hbar}{2 \rho V c_\mathrm{t} q}} \frac{d_\mathrm{p}e}{\epsilon_0\epsilon_\mathrm{r}} \sin(2\vartheta) \cos(2\varphi) \nonumber
\\
\mathcal{G}_\mathrm{TA2}(\mathbf{q}) =& -\i \sqrt{\frac{\hbar}{2 \rho V c_\mathrm{t} q}} \frac{d_\mathrm{p}e}{\epsilon_0\epsilon_\mathrm{r}} (3\cos^2 \vartheta - 1)\sin \vartheta \sin(2\varphi).
\end{align}
Here, $c_s$ is the speed of sound of the respective phonon branch $s$, $D$ is the deformation potential and $d_\mathrm{p}$ the piezoelectric constant. The $\mathbf{q}$-sum in Eq.~\eqref{eq:imunu} is replaced by the integral $\sum_\mathbf{q} \rightarrow V/(2\pi)^3 \int \d^3 \, q$ in the continuum limit and the acoustic phonon dispersion is assumed to be linear, i.e.~$\omega_ {\mathbf{q}, s} = c_s q$. With this, the spectral functions can be expressed as
\begin{equation}
I_{\mu\nu}^s (\omega) = \left.\int_0^\pi \d \vartheta\, \mathcal{K}_s(\omega, \vartheta)\mathcal{F}^\ast_\mu (\mathbf{q}) \mathcal{F}_\nu (\mathbf{q})\right\vert_{q=\omega/c_s}
\end{equation}
with 
\begin{equation}
\mathcal{K}_s(\omega, \vartheta) = \left.\frac{V\omega^2}{(2\pi)^3c_s^2} \sin\vartheta \int_0^{2\pi} \d \varphi \, |\mathcal{G}_s(\mathbf{q})|^2 \right\vert_{q=\omega/c_s}
\end{equation}
being determined by the coupling mechanism of the respective phonon branch.

\section{Transition to a Bloch-Redfield equation for an explicitly time-dependent system Hamiltonian}
In the derivation of the Bloch-Redfield equation, one encounters integrals of the form
\begin{equation}
\int_0^\infty \d \tau \, \braket{B_\alpha^\dag(t) B_\beta (t-\tau)} A_\beta(\omega, t-\tau) A^\dag_\alpha(\omega, t),
\end{equation}
where the operators are given in the interaction picture, see for example Ref.~\citep{breuer2002theory}. As long as the system Hamiltonian does not carry an explicit time dependence, the time dependence of the eigenoperators is given by
\begin{equation}
A_\alpha(\omega, t) = \e^{\i H_\mathrm{S}t/\hbar} A_\alpha (\omega)  \e^{-\i H_\mathrm{S}t/\hbar} = \e^{-\i\omega t}A_\alpha(\omega).
\end{equation}
After further simplifications, this can be written as
\begin{equation} \label{eq:D3}
\int_{-\infty}^\infty \d \tau \, \braket{B_\alpha^\dag(t) B_\beta (t-\tau)} \e^{\i \omega \tau} A_\beta(\omega) A^\dag_\alpha(\omega),
\end{equation}
leading to the definition of the spectral function discussed in the context of Eq.~\eqref{eq:replacement} in the main text. 

This is not directly possible in the time-dependent case, in which the eigenoperators  carry the non-trivial time dependence $A_\alpha(\omega, t) = U^\dag(t,t_0)A_\alpha(\omega,t_0) U(t,t_0)$, with the time evolution operator $U(t,t_0) = \mathrm{T} \exp \left(-\i/\hbar \int_{t_0}^t H_\mathrm{S}(t^\prime) \d t^\prime\right)$, T being the time-ordering operator \citep{fetter2012quantum}. The above form \eqref{eq:D3} can be recovered when only considering a small time window $\Delta t$, in which $H_\mathrm{S}(t)$ does not vary considerably. Since the integral is extended up to $\tau = \infty$, the correlation function $\braket{B_\alpha^\dag(t) B_\beta (t-\tau)}$, has to decay within $\Delta t$ to justify the replacement $A_\alpha(\omega, t) = \e^{-\i\omega t}A_\alpha(\omega)$ for a time-dependent Hamiltonian. 
\\
We can use the connection between the correlation function and and the spectral density  \citep{may2023charge}
\begin{equation}
C(\tau) = \int_0^\infty \d \omega\left(\cos (\omega\tau) \coth \frac{\hbar \omega}{2 k_\mathrm{B} T} - \i \sin (\omega \tau) \right)\omega^2 J (\omega)
\end{equation}
to investigate how the correlation function decays in time.
\begin{figure}[H]
	\centering
	\includegraphics[width=\columnwidth]{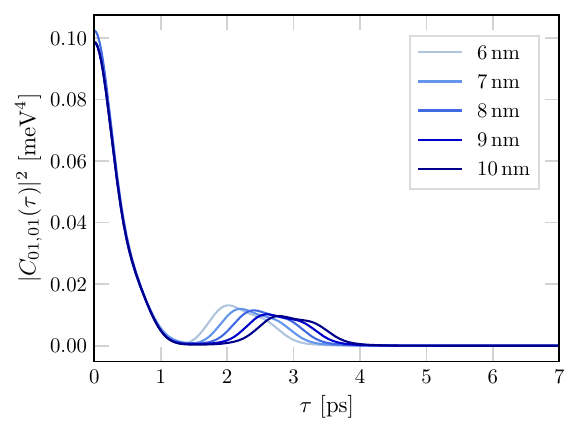}
	\caption{Correlation function $|C(\tau)|^2$ for different tunneling barrier widths $w$.}
	\label{fig:correlation_func}
\end{figure}
Figure~\ref{fig:correlation_func} shows the absolute value of this correlation function for different tunneling-barrier widths. Irrespective of $w$ (and therefore $t_\mathrm{e}$), the phonon correlation function decays quickly on the order of ps. This is faster than the switching times, for which we can expect adiabatic behavior.

\section{Parameters}
Finally, we list all parameters that have been used for the calculations of single-particle wave functions and phonon spectral functions in Table~\ref{tab:params}.
\begin{table}[h]
	\centering
	\begin{tabular}{m{5.2cm} m{3.1cm}}
	\hline 
		Parameter & Value \\ \hline \hline
		Oscillator length & $\beta_\mathrm{e}^{-1} = \SI{5.4}{nm}$ \\[1pt] 
		Potential depth & $E_\mathrm{d} = \SI{350}{meV}$ \\[1pt] 
		Heigth of quantum dots & $h = \SI{4.5}{nm}$ \\[1pt] 
		\hline \\[-8pt] 
		Effective mass & $m^\ast = 0.065 \, m_0$ \\[1pt] 
		Dielectric constant & $\epsilon_\mathrm{r} = 12.9$ \\[1pt] 
		Crystal density & $\rho = \SI{5300}{kg/m^3}$\\[1pt] 
		Longitudinal speed of sound & $c_\mathrm{l} = \SI{5.15}{ps/nm}$\\[1pt] 
		Transversal speed of sound & $c_\mathrm{t} = \SI{2.8}{ps/nm}$\\[1pt] 
		Deformation potential & $D = -\SI{6.66}{eV}$\\[1pt] 
		Piezoelectric coupling & $d_\mathrm{p} = -\SI{0.16}{C/m^2}$\\[1pt] 

		\hline
	\end{tabular}
\caption{QD geometry and material specific parameters, taken from \citep{gawarecki2010phonon} and used throughout this work.}\label{tab:params}
\end{table}

\newpage
\bibliography{references}
\end{document}